\documentclass[amsmath,amssymb,pra,twocolumn,showpacs]{revtex4}
\usepackage[dvips]{graphicx}
\usepackage{bm}

\begin{document}

\title{Statistics of voltage fluctuations in resistively shunted Josephson junctions}

\author{
D. S. Golubev$^1$,
M. Marthaler$^{2}$,
Y. Utsumi$^3$, and Gerd Sch\"on$^{1,2}$
}

\affiliation{
$^1$Karlsruhe Institute of Technology, Institut f\"ur Nanotechnologie, 76021 Karlsruhe, Germany \\
$^2$Institut f\"{u}r Theoretische Festk\"{o}rperphysik and
DFG Center for Functional Nanostructures (CFN)\\
Karlsruhe Institute of Technology, 76128 Karlsruhe, Germany \\
$^3$Institute for Solid State Physics, University of Tokyo, Kashiwa, Chiba 277-8581, Japan \\
}

\begin{abstract}
The intrinsic nonlinearity of Josephson junctions converts
Gaussian current noise in the input into non-Gaussian voltage noise in the output.
For a resistively shunted Josephson junction with white input noise
we determine numerically exactly the properties of
the few lowest cumulants of the voltage fluctuations, and we derive analytical
expressions for these cumulants in several important limits.
The statistics of the voltage fluctuations
is found to be Gaussian at bias currents well above the Josephson critical current, but
Poissonian at currents below the critical value. In the transition region close to the
critical current the higher-order cumulants oscillate and the voltage noise is
strongly non-Gaussian. For coloured input noise we determine
 the third cumulant of the voltage.
\end{abstract}


\maketitle

\section{Introduction}

Resistively shunted Josephson junctions, described in the frame of the
so-called RSJ model, have been widely studied for many years for the
analysis of the high frequency response, noise,
non-linear dynamics, and other properties of Josephson junctions \cite{Likharev,Barone}.
Analytical expressions for the current-voltage characteristics \cite{Ambegaokar}
and noise \cite{Likharev1972,Likharev}, as well as numerous numerical
results for this model are known.
In the present paper we analyse the statistics of the voltage
fluctuations in the RSJ model, which has not been addressed in the
literature before. Our work is partly motivated by recent progress with experimental techniques,
which made it possible to detect non-Gaussian corrections to the statistics
of current fluctuations in a normal tunnel junction \cite{Reulet2003,Timofeev2007,Peltonen2007,LeMasne2009}.
It should be possible to apply similar techniques to Josephson junctions.
Another motivation is to study the effects of
the intrinsic non-linearity inherent to the RSJ, which is absent
in normal tunnel junctions. One of its manifestations
are the Shapiro steps in the $I$-$V$ characteristic
of an $ac$-biased RSJ \cite{Likharev,Barone}.
The non-linearity also converts high-frequency input current noise
to low-frequency voltage noise. The latter effect allowed Koch {\it et al.}
to measure the high-frequency quantum noise of an Ohmic resistor \cite{Koch1982}.
In this paper we will study the effect of the Josephson non-linearity
on the statistics of voltage fluctuations, and we
find strong deviations of the latter from a simple Gaussian form.

We proceed using the formalism of the Full Counting Statistics (FCS) \cite{Levitov,Nazarov},
which is the appropriate tool for analysing non-Gaussian stochastic processes.
The concept of FCS proved to be very
useful in the transport theory of mesoscopic systems. So far most efforts
focused on the FCS of electron transport through tunnel junctions,
quantum dots, metallic wires and similar
mesoscopic structures \cite{Nazarov,Bagrets}.
The first two cumulants - current and zero-frequency current noise -
are routinely measured in experiments. Higher cumulants provide information about deviations from
Gaussian statistics, and are therefore of great importance.
The FCS formalism has also been applied to the charge transport through
a voltage biased Josephson junction \cite{Cuevas2003,Johansson2003,Cuevas2004},
where it was shown that the statistics of current fluctuations at bias
voltages below the superconducting energy gap is determined by
the processes of multiple Andreev reflections.

On the experimental side, a direct measurement of the third cumulant
for a tunnel junction between two normal metals
has been first performed by Reulet {\it et al.} \cite{Reulet2003}
and good agreement with theory \cite{Beenakker2003} was found.
Subsequently it has been realized \cite{Tobiska2004,Pekola2004}
that a current biased Josephson junction can be used as threshold detector of large
current fluctuations,
which makes it an effective tool of measuring higher order cumulants
of non-Gaussian noise. This idea has been implemented in experiments
\cite{Timofeev2007,Peltonen2007,LeMasne2009} where the third cumulant of the
current noise of a tunnel junction was measured.
In several studies \cite{Grabert2008,Sukhorukov2007,Ankerhold}
the theory of a Josephson junction, serving as detector of
non-Gaussian noise, has been developed in detail.

In this paper we treat the resistively shunted Josephson junction as a source,
rather than a detector, of non-Gaussian noise.
In contrast to Refs. \cite{Cuevas2003,Johansson2003,Cuevas2004} we consider
a current biased junction and study the statistics of the resulting voltage fluctuations.
After briefly introducing the
RSJ model of an overdamped Josephson junction
and the concept of  FCS for voltage noise,
we will study the FCS of voltage fluctuations caused
by white Gaussian input current noise.
Instead of counting transferred electrons, we study the statistical distribution
of the Josephson phase.
We find that the statistics of voltage fluctuations
in a RSJ is Gaussian at bias currents far above
the critical current $I \gg I_c$, and Poissonian in the
opposite limit, $I\lesssim I_c$. In the transition region, $I\sim I_c$,
the voltage cumulants oscillate, and
at low temperature the statistics may become strongly non-Gaussian.
In the last part of the paper we turn to the case of colored input noise
and derive approximate expressions for the third cumulant of the voltage noise.

\section{Model}

An overdamped resistively shunted Josephson junction is described by the equation
\cite{Likharev,Barone}
\begin{eqnarray}
\frac{1}{R}\frac{\hbar\dot\varphi}{2e}+I_c\sin\varphi = I+\xi(t).
\label{RSJ}
\end{eqnarray}
Here $\varphi$ is the phase difference of the superconducting order
parameters across the junction, $R$ is the parallel shunt resistance,
$I_c$ and $I$ are the critical current and bias current due to a current source, respectively,
and $\xi(t)$ is the input Gaussian noise.
In equilibrium $\xi(t)$ is just the noise of the resistance $R$ fixed
by the fluctuation-dissipation theorem,
\begin{eqnarray}
\langle \xi(t_1)\xi(t_2) \rangle = \int\frac{d\omega}{2\pi}e^{-i\omega(t_1-t_2)}
\frac{\hbar\omega}{R}\coth\frac{\hbar\omega}{2T} \, .
\label{fdt}
\end{eqnarray}
At high temperature, $T\gtrsim 2eI_cR$,
the equilibrium noise $\xi(t)$ is white with correlator
\begin{eqnarray}
\langle \xi(t_1)\xi(t_2) \rangle = \frac{2T}{R}\, \delta(t_1-t_2) \, .
\label{white}
\end{eqnarray}
Out of equilibrium the noise spectrum may be arbitrary, but often
it is sufficiently well described by Eq. (\ref{white}) with an
enhanced effective temperature $T^*$. At low frequencies,
$1/f$ noise usually dominates, but we will not consider it here.
In Sec. IV we will also consider coloured noise with
an arbitrary spectrum requiring only that it does not diverge
in the limit of low frequencies.

We note that Eq.~(\ref{RSJ}) is not exact since it ignores charge accumulation
on the capacitance $C$ of the junction and the quantum dynamics derived from this effect.
At high resistance, namely for $R\gtrsim \pi\hbar/2e^2$,
this effect is  not important at sufficiently high temperatures,
$T\gtrsim \sqrt{2 \hbar eI_c /C}$.
In the opposite limit $R\ll \pi\hbar/2e^2$ quantum effects may be ignored at any
temperature (see e.g. Ref. [\onlinecite{Schoen}] for more details).

For single-electron tunneling in normal junctions
the FCS is contained in the probability $W(t,N)$ for $N$ electrons
to be transfered through the system during time $t$.
The central object of FCS theory is the cumulant generating function
\begin{eqnarray}
S(\chi)=\ln\left[\sum_N e^{iN\chi}W(t,N)\right],
\label{S}
\end{eqnarray}
where $\chi$ is the so-called counting field. Derivatives of this function
determine the zero-frequency cumulants of the electric current flowing through the system,
\begin{eqnarray}
C_n^{(I)}=\lim_{t\to \infty}\frac{1}{t}\frac{1}{i^n}\frac{\partial^n S(\chi)}{\partial \chi^n}\bigg|_{\chi=0}.
\end{eqnarray}
Proceeding in analogy to standard FCS theory of electron
transport, we first introduce for the Josephson junction the probability $P(t,\varphi)$
for the Josephson phase to change by $\varphi$ during time $t$.
The corresponding cumulant generating function reads
\begin{eqnarray}
F(t,k)       &=&\ln\left[{\cal P}\right],\\
{\cal P}(t,k)&=&\int d\varphi\, e^{ik\varphi}P(t,\varphi)\, ,\label{Pk}
\end{eqnarray}
where we have introduced the counting field $k$
and the Fourier transformed distribution function ${\cal P}(t,k)$.
The zero-frequency cumulants of the voltage $V=\hbar\dot\varphi/2e$ are given by the
derivatives
\begin{eqnarray}\label{eq:Cumulants}
C_n^{(V)}= \frac{\hbar^n}{(2e)^n}\lim_{t\to\infty}\frac{(-i)^n}{t}\frac{\partial^n F(t,k)}{\partial k^n}\bigg|_{k=0}.
\label{CV}
\end{eqnarray}
Note that in contrast to
the usual cumulant generating function (\ref{S})
the function $F(t,k)$ is not periodic in $k$.
The reason for that is the fact that the transferred charge is discrete
while the phase $\varphi$ is continuous.

To proceed we rewrite Eq. (\ref{RSJ}) in dimensionless form
by defining the dimensionless parameters
$\tau=2eI_cRt/\hbar$, $j=I/I_c$, $\zeta=\xi/I_c$,
and obtain
\begin{eqnarray}
\frac{d\varphi}{d\tau}+\sin\varphi=j+\zeta(\tau).
\label{RSJ1}
\end{eqnarray}
The correlator of the dimensionless noise $\zeta(\tau)$
reads
\begin{eqnarray}
\langle \zeta(\tau_1)\zeta(\tau_2)\rangle = \int\frac{d\nu}{2\pi}\,e^{-i\nu(\tau_1-\tau_2)}\,S(\nu),
\end{eqnarray}
where $\nu = \hbar\omega/2eI_cR$ is the dimensionless frequency, and
\begin{eqnarray}
S(\nu) = \frac{2eR}{\hbar I_c}\int dt\, \exp\left(i\frac{2eI_cR\nu }{\hbar}t\right)\, \langle\xi(t)\xi(0)\rangle
\end{eqnarray}
is the dimensionless noise spectrum.
In the same way we normalize the cumulants.
The zero frequency $n^{\rm th}$ cumulant of the dimensionless voltage $v=d\varphi/d\tau$ becomes
\begin{eqnarray}\label{eq:CumulantsNormalized}
{\cal C}_n= \lim_{\tau\to\infty}\frac{(-i)^n}{\tau}\frac{\partial^n F(\tau,k)}{\partial k^n}.
\end{eqnarray}
The cumulant (\ref{CV}), with dimensions restored, is then given by  $C_n^{(V)}=(I_cR)^n{\cal C}_n$.

The two lowest cumulants
are directly related to measurable physical quantities. The first  is
 the average voltage $ {\cal C}_1 \equiv \langle v\rangle$,
while the second  is the zero-frequency voltage
noise ${\cal C}_2 \equiv \int d\tau \langle \delta v(\tau)\delta v(0)\rangle$,
with $\delta v(\tau)=v(\tau)-\langle v\rangle$. In the following we will also use
the explicit expression for the zero-frequency third cumulant
\begin{eqnarray}
{\cal C}_3\equiv \int d\tau_1d\tau_2  \langle \delta v(\tau_1)\delta v(\tau_2)\delta v(0) \rangle.
\label{C3def}
\end{eqnarray}

\section{White noise}

In this section we consider white noise with
local pair correlator (\ref{white}).
In this regime the Langevin equation (\ref{RSJ1}) is equivalent to the
Smoluchowski equation for the distribution function $P(\tau,\varphi)$,
\begin{eqnarray} \label{eq:RSJ}
\frac{\partial P}{\partial\tau}+
\frac{\partial}{\partial \varphi}\big(  (j-\sin\varphi)\; P \big) - \gamma
\frac{\partial^2 P }{\partial \varphi^2}  =0.
\label{smoluchowski}
\end{eqnarray}
Here we  introduced the dimensionless constant $\gamma = 2e T/\hbar I_c$
characterizing the noise strength.

The average voltage ${\cal C}_1$ in this case is known exactly
and reads \cite{Ambegaokar}
\begin{eqnarray}
{\cal C}_1=\frac{2\pi\gamma}
{ \int_{-\pi}^{\pi} d\varphi_1\int_{-\infty}^{\varphi_1}d\varphi_2\;
e^{[j(\varphi_1-\varphi_2)+\cos\varphi_1-\cos\varphi_2]/\gamma} }.
\label{C1exact}
\end{eqnarray}
In order to access higher order cumulants
we have solved Eq.~(\ref{smoluchowski}) numerically and, in some limiting cases,
analytically. Let us first present the numerical solution.
The Fourier transformed distribution
function ${\cal P}(t,k)$
satisfies the following equation
\begin{eqnarray}
\frac{\partial {\cal P}(t,k)}{\partial\tau} &=&
ik\left(  j {\cal P}(t,k) -\frac{{\cal P}(t,k+1)-{\cal P}(t,k-1)}{2i} \right)\nonumber\\
& &- \gamma k^2{\cal P}(t,k).
\end{eqnarray}
In order to solve this equation we define a square matrix $\hat L(k)$ with  elements
\begin{eqnarray}
L_{\alpha\beta}(k)&=&\big[ij(k+\alpha)-\gamma (k+\alpha)^2\big]\delta_{\alpha,\beta}
\nonumber\\ &&
-\,\frac{k+\alpha}{2}(\delta_{\alpha,\beta-1}-\delta_{\alpha,\beta+1})\, ,
\label{L}
\end{eqnarray}
and truncate it restricting the indeces $\alpha$ and $\beta$ as follows: $-N/2 \leq \alpha,\beta \leq N/2$, where $N$ is sufficiently large. For the results presented below we
 have chosen $N=400$, which guarantees sufficient accuracy.
Next we note that the long-time behavior of $F(\tau,k)$
is given by a simple formula
\begin{eqnarray}
F(\tau,k)=\lim_{\tau\to\infty}\ln {\cal P}(\tau,k)= \tau\Lambda(k),
\label{F}
\end{eqnarray}
where $\Lambda(k)$ is the eigenvalue of the matrix $\hat L(k)$ with
the biggest real part.
The numerically exact results for the first three cumulants obtained in this way
are shown in Figs. 1-4.

Before discussing these results further we first consider in the following section
 the limits of high and low temperatures, where we can derive analytical results.

\subsection{High temperature regime, $\gamma \gtrsim 1$}

\begin{figure}
\includegraphics[width=0.8 \columnwidth]{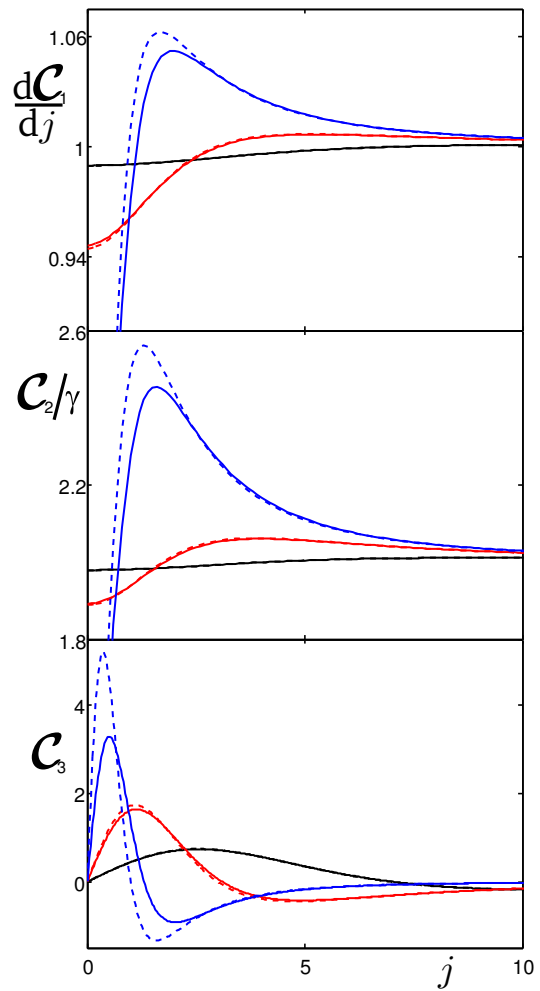}
\caption{The derivative of the normalized voltage $d{\cal C}_1/dj$, the second cumulant
normalized by the temperature ${\cal C}_2/\gamma$ and the third cumulant ${\cal C}_3$
as a function of the bias current $j$. (black) $\gamma=7$, (red) $\gamma=3$,
(blue) $\gamma=1$. (full lines) numerical results,
(dashed line) analytical results (\ref{Chigh}).
}
\label{fig:C1C2C3largega}
\end{figure}

In the limit of high temperature, $\gamma\gtrsim 1$, we use a simple perturbation theory
which allows us to find an analytical expression for the cumulant generating function (\ref{F})
at any bias current.
Treating the off-diagonal elements of the matrix
(\ref{L}) as weak perturbation we find the lowest order correction
to the dominating eigenvalue $\Lambda(k)$. This simple procedure relies on
perturbation theory in the critical current $I_c$ in the original
RSJ equation (\ref{RSJ}). For this approximation to be justified, $I_c$
should be small compared to bias current $I$ or some scale set by the temperature $T$.
More precisely, in terms of dimensionless parameters this requirement can be written
as $\sqrt{j^2+\gamma^2}\gtrsim 1$.
Thus we arrive at the following expression for the cumulant generating function
\begin{eqnarray}\label{eq:perturbation}
F(\tau,k)=\bigg(ijk-\gamma k^2 + \frac{1}{2}\frac{ijk-\gamma k^2}{(ij-2\gamma k)^2-\gamma^2} \bigg)\tau.
\label{highT}
\end{eqnarray}
The first three cumulants read
\begin{eqnarray}
{\cal C}_1 &=& j\left(1-\frac{1}{2(j^2+\gamma^2)}\right),
\nonumber\\
{\cal C}_2 &=& \gamma \frac{2j^4+3j^2+4j^2\gamma^2+2\gamma^4-\gamma^2}{(j^2+\gamma^2)^2},
\nonumber\\
{\cal C}_3 &=& 24 \gamma^2\, j\,\frac{\gamma^2-j^2}{(j^2+\gamma^2)^3}.
\label{Chigh}
\end{eqnarray}
We observe that the voltage fluctuations become
purely Gaussian (${\cal C}_{n}\to 0$, $n\geq 3$) at $\gamma\rightarrow \infty$ or $j\to\infty$
because in this regime the nonlinear Josephson current may be ignored.

In Fig. \ref{fig:C1C2C3largega} we compare analytical results (\ref{Chigh})
with the exact numerical simulations.
While analytics and numerics agree well for the first and second cumulant, even for rather
small $\gamma$, the analytical approximation for the third cumulant breaks down for
$\gamma \approx 1$.
The most interesting observation is the sign change of ${\cal C}_3$.
It is positive at low bias, negative at high bias and equals zero
at $j=\gamma$. According to Eq. (\ref{highT}) the
higher order cumulants oscillate even stronger: ${\cal C}_4,{\cal C}_5$
change their sign twice, ${\cal C}_6,{\cal C}_7$ -- three times,
${\cal C}_8,{\cal C}_9$ -- four times, etc. An exact numerical solution
confirms this fact
and shows that the oscillations of the high order
cumulants occur even at $\gamma\ll 1$, where the Eq. (\ref{highT})
is no longer valid (see Fig. \ref{fig:C3C4C5smallga}).
As it has been
realized recently \cite{flindt}, such an oscillatory behavior of
the cumulants is a very general phenomenon.

The third cumulant ${\cal C}_3$ (\ref{Chigh}) has two extrema at $j=\gamma(4\pm\sqrt{13})^{1/2}/\sqrt{3}$.
The extremal values of ${\cal C}_3$ are proportional to
$1/\gamma$, $C_3^{\rm max/min}\propto 1/\gamma $. As $\gamma$
becomes smaller, ${\cal C}_3$ increases while ${\cal C}_2\propto\gamma$ decreases.
At $\gamma\sim 1$, where the approximation (\ref{highT}) breaks down,
we have ${\cal C}_2\sim {\cal C}_3$.
This result indicates significant deviation
from the simple Gaussian statistics of voltage fluctuations.
We will explore this region now coming from the opposite limit
of small $\gamma$.

\subsection{Low temperature regime, $\gamma\lesssim 1$}

\begin{figure}[t]
\includegraphics[width=0.8 \columnwidth]{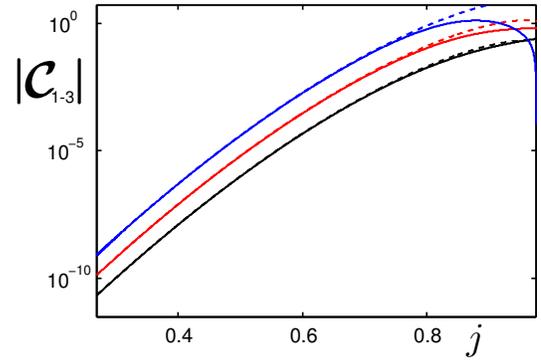}
\caption{The absolute value of the first three Cumulants $|{\cal C}_{1-3}|$,
as a function the bias  current $j$ for $j<1$ and $\gamma=0.05$.
(black) the first Cumulant ${\cal C}_1$, (red) the second Cumulant ${\cal C}_{2}$,
(blue) the third cumulant ${\cal C}_3$. (full lines) numerical results, (dashed lines)
analytical results (see eq. (\ref{Poisson}))
}
\label{fig:C1C2C3smallga1}
\end{figure}

At low temperature, $\gamma\lesssim 1$,
we will distinguish low bias regime, $(1-j)^{3/2}\gtrsim\gamma$, and
high bias regime, $(j-1)^{3/2}\gtrsim\gamma$.
In the vicinity of the critical current, namely at $|j-1|^{3/2}\lesssim \gamma$, we
restrict ourselves to numerics.

Let us first consider low bias currents $(1-j)^{3/2}\gtrsim\gamma$.
In this case the phase is strongly
localized at the minima $\varphi_n=\arcsin j+2\pi n$
of the effective potential
$U(\varphi)=-j\varphi-\cos\varphi$ \cite{Haenggi1990}.
Defining the occupation probability $P_n(\tau)$ of the $n$-th minimum
one can express the function $P(\tau,\varphi)$ in the form
\begin{eqnarray}\label{eq:low_temperatures_small_j_approx_p}
P(\tau,\varphi)=\sum_n P_n(\tau)
\frac{(1-j^2)^{1/4}e^{-(\varphi-\varphi_n)^2\sqrt{1-j^2}/2\gamma}}{\sqrt{2\pi\gamma}}.
\end{eqnarray}
$P_n(\tau)$ satisfies the following master equation \cite{Haenggi1990}
\begin{eqnarray} \label{balance}
\frac{\partial P}{\partial\tau}=-(\Gamma_+ +\Gamma_-)P_n+\Gamma_+P_{n-1}+\Gamma_-P_{n+1},
\end{eqnarray}
where
\begin{eqnarray}
\Gamma_\pm=\frac{\sqrt{1-j^2}}
{2\pi}\exp\left[-2\frac{j\arcsin j+\sqrt{1-j^2}}{\gamma} \pm\frac{\pi j}{\gamma}\right].
\end{eqnarray}
The Fourier transformed distribution function ${\cal P}(\tau, k)$ in this case acquires the form
\begin{eqnarray}
{\cal P}(\tau,k) = e^{i\arcsin j} e^{-\gamma k^2/2\sqrt{j^2-1}}\sum_n e^{2\pi i kn}P_n(\tau)
\end{eqnarray}
and satisfies the equation
\begin{eqnarray}
\frac{\partial {\cal P}}{\partial\tau}=
\left[\Gamma_+\big( e^{2\pi ik}-1 \big)+\Gamma_-\big( e^{-2\pi ik}-1 \big)\right]{\cal P}.
\end{eqnarray}
Thus the cumulant generating function takes the form
\begin{eqnarray}\label{Poisson}
F(\tau,k)=\Gamma_+\tau\big( e^{2\pi ik}-1 \big)+\Gamma_-\tau\big( e^{-2\pi ik}-1 \big).
\end{eqnarray}
The first three cumulants in this regime read
\begin{eqnarray}
{\cal C}_1 &=& 2\sqrt{1-j^2}\sinh(\pi j/\gamma)e^{-2\left(j\arcsin j +\sqrt{1-j^2} \right)/\gamma},\nonumber\\
{\cal C}_2 &=& 2\pi \coth (\pi j/\gamma)\,{\cal C}_1,\nonumber\\
{\cal C}_3 &=& 4\pi^2 \,{\cal C}_1.
\end{eqnarray}

The cumulant generating function (\ref{Poisson}) describes the Poissonian statistics
of bidirectional phase slip tunneling.
The voltage fluctuations in this case are determined by random
switches between the wells
of the tilted washboard potential. In Fig. \ref{fig:C1C2C3smallga1}
we compare analytical results with numerical ones and find an excellent agreement between both
as long as $(1-j)^{3/2}\gtrsim \gamma$, i.e. as long as the height of
the potential barrier between the neighboring minima is greater than
the temperature. At low currents the tunneling rate between the minima
becomes exponentially small and all cumulants are strongly
suppressed.

Next we assume that the bias current is bigger than the critical one
but not too close to it, i.e. $j>1$ and $(j-1)^{3/2}\gtrsim\gamma$.
To find the eigenvalue $\Lambda$ of the operator $L$ (\ref{L}) we rewrite it back in phase
representation and arrive at the following equation
\begin{eqnarray}
-\frac{\partial}{\partial\varphi}\left((j-\sin\varphi) \Psi\right)+\gamma\frac{\partial^2\Psi}{\partial\varphi^2}=\Lambda\Psi,
\label{eigen}
\end{eqnarray}
where the eigenfunction $\Psi$ is quasiperiodic: $\Psi(\varphi+2\pi)=\exp[-2\pi i k]\Psi(\varphi)$.
Our aim is to find the eigenvalue $\Lambda$ as a function of the counting field $k$.
We express the function $\Psi$ as follows
\begin{eqnarray}
\Psi(\varphi)=\exp\left[\frac{1}{\gamma}\int_0^\varphi d\varphi'\,W(k,\varphi')\right].
\end{eqnarray}
Eq. (\ref{eigen}) then takes the form
\begin{eqnarray}
\gamma\frac{\partial W}{\partial\varphi}=(j-\sin\varphi)W-W^2-\gamma\cos\varphi +\gamma\Lambda.
\label{W}
\end{eqnarray}
The quasiperiodicity of the function $\Psi$ translates into the following condition
\begin{eqnarray}
\frac{1}{2\pi}\int_0^{2\pi} d\varphi\, W(\varphi)=-i\gamma k.
\label{Wk}
\end{eqnarray}

\begin{figure}[t]
\begin{center}
\includegraphics[width=0.8 \columnwidth]{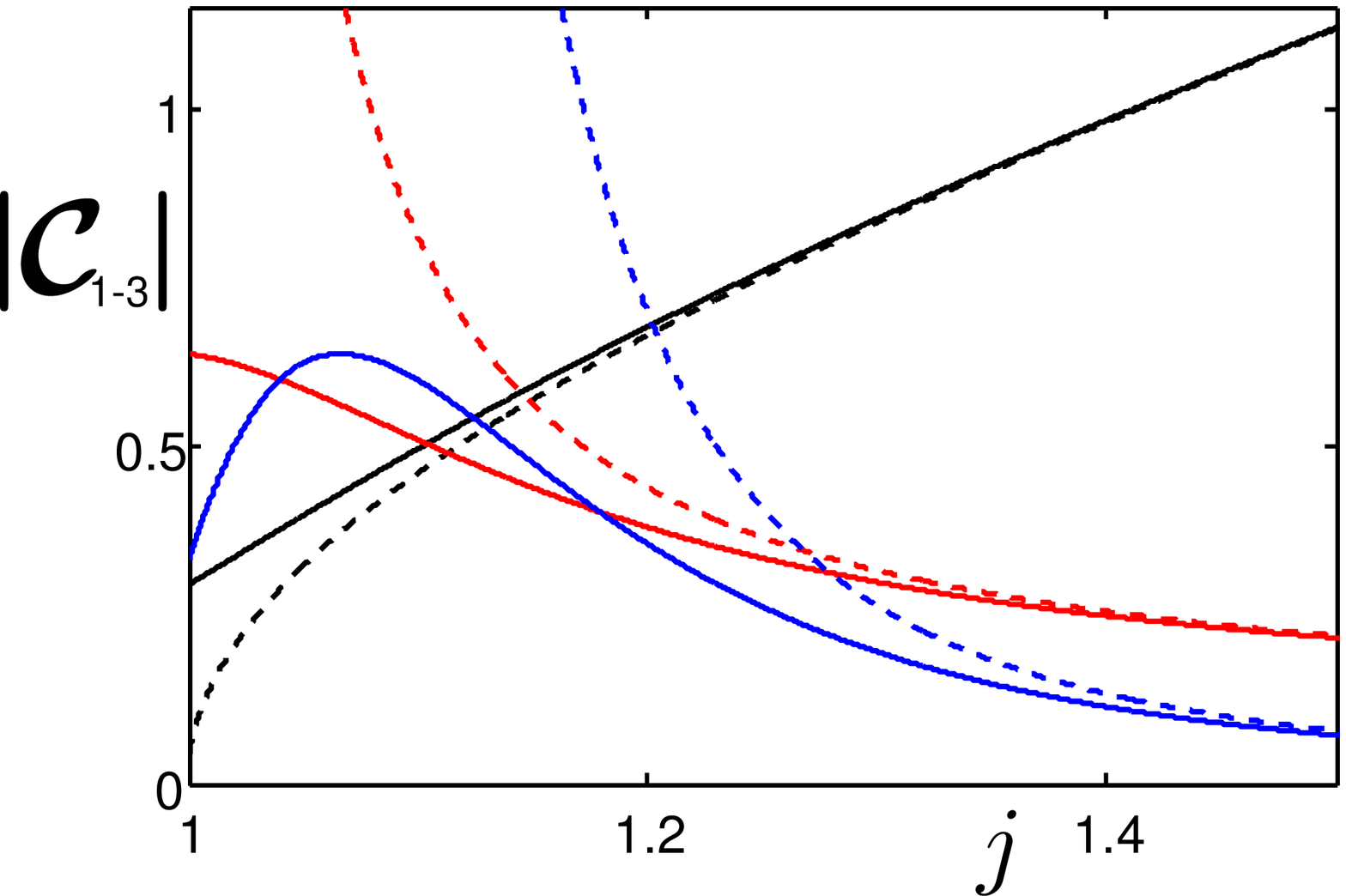}
\caption{The absolute value of the first three Cumulants $|{\cal C}_{1-3}|$,
as a function the bias  current $j$ for $j>1$ and $\gamma=0.05$.
(black) the first Cumulant ${\cal C}_1$, (red) the second Cumulant ${\cal C}_{2}$,
(blue) the third cumulant ${\cal C}_3$. (full lines) numerical results, (dashed lines)
analytical results (see Eqs.. (\ref{vax}), (\ref{likharev}) and (\ref{C3white})).
}
\label{fig:C1C2C3smallga2}
\end{center}
\end{figure}

In the limit $\gamma\ll 1$ and $(j-1)^{3/2}\gtrsim \gamma$, which
we are considering,
the solution of Eq. (\ref{W}) reads
\begin{eqnarray}
W(\varphi)=\frac{j-\sin\varphi-\sqrt{(j-\sin\varphi)^2+4\gamma\Lambda}}{2}.
\end{eqnarray}
Combining this result with Eq. (\ref{Wk}) we find the eigenvalue $\Lambda$ and obtain the
cumulant generating function in the form
\begin{eqnarray}
F(\tau,k)=\frac{f(j,2i\gamma k)}{4\gamma}\tau,
\label{F2}
\end{eqnarray}
where the function $f(j,x)$ is implicitly defined by the equation
\begin{eqnarray}
-j + \frac{1}{2\pi}\int_0^{2\pi}d\phi\,\sqrt{f(j,x)+(j-\sin\phi)^2}=x.
\label{f}
\end{eqnarray}
Having derived the cumulant generating function (\ref{F2},\ref{f}) we
can find the first three cumulants:
\begin{eqnarray}
{\cal C}_1 &=& \sqrt{j^2-1},
\label{vax}\\
{\cal C}_2 &=& \frac{2j^2+1}{j^2-1}\,\gamma,
\label{likharev} \\
{\cal C}_3 &=& -\frac{3}{2}\frac{16j^2+1}{(j^2-1)^{5/2}}\,\gamma^2.
\label{C3white}
\end{eqnarray}
The first cumulant (\ref{vax}) is just the usual voltage-current dependence
of an RSJ junction \cite{Likharev,Barone}, while the second one (\ref{likharev}) follows from
the well known result by Likharev and Semenov \cite{Likharev1972,Likharev}.

\begin{figure}[t]
\begin{center}
\includegraphics[width=0.8 \columnwidth]{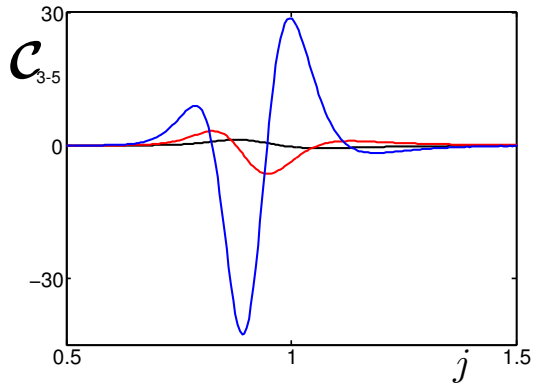}
\caption{Numerical solutions for the third, fourth and fifth cumulant ${\cal C}_{3-5}$,
as a function the bias  current $j$ for $\gamma=0.05$.
(black) the third Cumulant ${\cal C}_3$, (red) the fourth Cumulant ${\cal C}_{4}$,
(blue) the fifth cumulant ${\cal C}_5$.
}
\label{fig:C3C4C5smallga}
\end{center}
\end{figure}

In the vicinity of the critical current the $n$-th cumulant can be roughly estimated as follows:
\begin{eqnarray}
{\cal C}_n\propto (-1)^n\sqrt{j-1}\left[\frac{\gamma}{(j-1)^{3/2}}\right]^{n-1}.
\label{estimate}
\end{eqnarray}
The divergence close to $j=1$ is due to the fact that our approach is only valid
for $(j-1)^{3/2}\gtrsim \gamma$. To get a rough estimate of the cumulants
at $(j-1)^{3/2} < \gamma$ one should replace $(j-1)^{3/2}$ by $\gamma$
in Eq. (\ref{estimate}), which leads to the result $|{\cal C}_n|\sim \alpha^n {\cal C}_1$, where $\alpha >1 $ is
some constant.
This relation indicates gradual transition from Gaussian statistics observed at $(j^2-1)^{3/2}\gg \gamma$
to the Poissonian statistics (\ref{Poisson}) valid below the critical current.

The full numerical solution for the first three cumulants
shown in Fig. \ref{fig:C1C2C3smallga2}. It demonstrates that
the approximate analytical results (\ref{vax}-\ref{C3white})
work well at sufficiently high bias current. We also observe
that close to the critical current the analytics is less accurate
for the third cumulant.

In the transition region $|j-1|^{3/2} \lesssim \gamma$ the higher order cumulants oscillate as illustrated in Fig. \ref{fig:C3C4C5smallga}.
The amplitude of these oscillations increases with the order of the cumulant.
This shows that an overdamped Josephson junction at low temperatures
may be a very good source of non-Gaussian noise. Moreover, since the cumulants
oscillate out of phase, one can tune the ratio between them
varying the bias current. We observe, for example, that ${\cal C}_4=0$
at the bias current where $|{\cal C}_3|$ and $|{\cal C}_5|$ are close to their
maximum values.

\section{Coloured noise}

In this section we consider the effect of frequency dependent, or coloured noise.
It is well known that
due to the non-linearity of an RSJ the high-frequency part of the current noise $\xi(t)$
contributes to low-frequency voltage noise ${\cal C}_2$ \cite{Likharev1972,Koch1982}.
In this section we show that the third cumulant ${\cal C}_3$ is sensitive
to the high-frequency current noise as well.
Since the Smoluchowski equation (\ref{smoluchowski}) in no longer valid, we apply
perturbation theory in the Josephson critical current $I_c$ and current noise $\xi(t)$
deriving analytical expressions for ${\cal C}_3$ valid in the limits of strong
and weak current noise. As an application of our general results,
we study the behavior of the third cumulant in the vicinity
of a Shapiro step.

\subsection{Strong noise}

\begin{figure}
\includegraphics[width=0.8 \columnwidth]{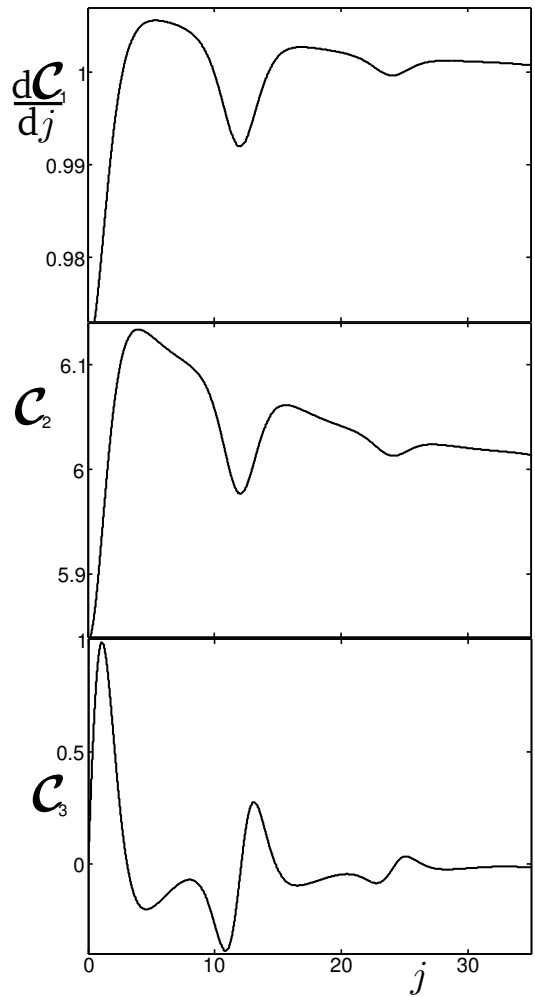}
\caption{The normalized differential resistance $\partial{\cal C}_1/\partial j$, the second cumulant (noise)
 ${\cal C}_2$ and the third cumulant ${\cal C}_3$ of an RSJ
subject to strong microwave irradiation as a function of the bias current $j$. ${\cal C}_1,{\cal C}_2,{\cal C}_3$
are defined by Eqs. (\ref{C1strong}),(\ref{C2strong}),(\ref{C3strong}) respectively,
while the noise spectrum -- by Eq. (\ref{AC}).
We have chosen the following values for the parameters characterizing the noise spectrum:
$\gamma=3$, $a=100$, $\delta=0.01$ and $\omega_0=12$.
}
\label{fig:C1C2C3lcolored}
\end{figure}

In case of strong noise we can apply the perturbation theory in critical current
$I_c$ while solving Eq. (\ref{RSJ}). Keeping the terms $\propto I_c^2$ and
switching to the dimensionless parameters, from Eq. (\ref{RSJ1}) we find
\begin{eqnarray}
v(\tau)\equiv\dot\varphi(\tau)=j+\zeta(\tau) - \sin\left[j\tau+\int_0^\tau ds\,\zeta(s)\right]
\hspace{1.2cm}
\nonumber\\
+\,\cos\left[j\tau+\int_0^\tau ds\,\zeta(s)\right]
\int_0^\tau ds_1 \sin\left[js_1+\int_0^{s_1} ds\,\zeta(s)\right].
\nonumber
\end{eqnarray}
The first cumulant is given by the average of this expression over the
fluctuations of the Gaussian noise $\zeta(\tau)$ and has
a well known form
\begin{eqnarray}
{\cal C}_1= j-\frac{1}{2}\int_0^\infty d\tau\, e^{-{\cal F}(\tau)}\sin j\tau,
\label{C1strong}
\end{eqnarray}
where
\begin{eqnarray}
{\cal F}(\tau)=\int \frac{d\nu}{2\pi}S(\nu)\frac{1-\cos\nu\tau}{\nu^2}.
\end{eqnarray}
The second cumulant in this regime reads
\begin{eqnarray}
{\cal C}_2=S(0)+\int_0^\infty d\tau\,\big[1-S(0)\tau\big]\,e^{-{\cal F}(\tau)}\cos j\tau.
\label{C2strong}
\end{eqnarray}
This expression can also be written in the form
\begin{eqnarray}
{\cal C}_2=S(0)\left( 2\frac{\partial{\cal C}_1}{\partial j}-1 \right)
+\frac{1}{2}\int d\tau\,e^{-{\cal F}(\tau)}\cos j\tau.
\end{eqnarray}

The third cumulant, which is of primary interest for us here, is defined
by Eq. (\ref{C3def}). After averaging of the latter over noise
fluctuations, we arrive at the following result
\begin{eqnarray}
{\cal C}_3 &=& \frac{3}{2}S^2(0)\int_0^\infty d\tau\, \tau^2\,e^{-{\cal F}(\tau)}\sin j\tau
\nonumber\\ &&
-\,3S(0)\int_0^\infty d\tau\, \tau\,e^{-{\cal F}(\tau)}\sin j\tau.
\label{C3strong}
\end{eqnarray}
One can verify that in case of white noise, $S(\nu)=2\gamma$, Eqs. (\ref{C1strong},\ref{C2strong},\ref{C3strong})
reduce to expressions (\ref{Chigh}) derived in Sec. IIIA.
In the regime of strong noise, considered here, the third cumulant can be related the first and second ones
as follows:
\begin{eqnarray}
{\cal C}_3 = 3S(0) \left(\frac{\partial{\cal C}_2}{\partial j} - S(0)\frac{\partial^2{\cal C}_1}{\partial j^2}\right).
\label{identity}
\end{eqnarray}

The results (\ref{C1strong},\ref{C2strong},\ref{C3strong}) are valid if
the correction to differential resistance at zero bias remains small, i.e. if
\begin{eqnarray}
\int_0^\infty d\tau\, e^{-{\cal F}(\tau)} \tau \ll 1.
\end{eqnarray}

Let us now investigate the behavior of the third cumulant in the vicinity
of a Shapiro step. Shapiro steps appear on the voltage-current characteristics
of an RSJ subject to an $ac$-bias \cite{Likharev,Barone}. In this case the noise $\xi(t)$
in Eq. (\ref{RSJ}) is the sum of the white noise and an $ac$ signal with the frequency $\omega_0$,
the line-width $\delta$ and the amplitude $\sqrt{a}$.
The corresponding noise spectrum reads
\begin{eqnarray}
S(\nu)=2\gamma + \frac{a\delta}{(\nu-\omega_0)^2+\delta^2}+\frac{a\delta}{(\nu+\omega_0)^2+\delta^2},
\label{AC}
\end{eqnarray}
and the function ${\cal F}(\tau)$ takes the form
\begin{eqnarray}
{\cal F}(\tau)&=&\gamma|\tau|
+a\frac{(\omega_0^2-\delta^2)(1-e^{-\delta|\tau|}\cos[\omega_0\tau])}{(\omega_0^2+\delta^2)^2}
\nonumber\\ &&
-\, a\frac{2\delta\omega_0 e^{-\delta|\tau|}\sin[\omega_0|\tau|]}{(\omega_0^2+\delta^2)^2}.
\label{F1}
\end{eqnarray}

Fig.~\ref{fig:C1C2C3lcolored} shows the behavior of the first three cumulants
for an RSJ subject to a strong Gaussian noise with the spectrum (\ref{AC}).
In the strong noise limit considered here the Shapiro steps occur at bias currents
$j_n=n\omega_0$. They are seen as dips in differential resistance $\partial{\cal C}_1/\partial j$.
We observe that the second cumulant ${\cal C}_2$ roughly follows the differential
resistance, while ${\cal C}_3$ exhibits beatings and changes its sign in the vicinity of every
Shapiro step.

\subsection{Weak Noise}

In this subsection we assume the noise $\zeta(\tau)$ to be weak.
More precisely, we require
\begin{eqnarray}
S(\nu_0)\ll \nu_0^2,
\end{eqnarray}
where $\nu_0=\sqrt{j^2-1}$ is the frequency of Josephson
oscillations.

Below the critical current, i.e. at $j<1$, we expect the Poissonian
statistics of the voltage noise  defined by Eq. (\ref{Poisson})
with modified tunneling rates $\Gamma_{\pm}$. It is in general not possible
to derive a closed formula for these rates, although for certain noise
spectra $S(\nu)$ it can be done \cite{Haenggi1990,Pekola2005}.

Above the critical current, i.e. for $j>1$, we slove Eq.
(\ref{RSJ1}) treating the noise $\zeta(\tau)$ perturbatively. The
first step of our approach is to put $\zeta(\tau)=0$ in Eq.
(\ref{RSJ1}) and find the zeroth order solution of this equation.
It can be expressed as follows
\begin{eqnarray}
\varphi_0(\tau)=\nu_0\tau + \Delta\varphi_0(\tau),
\end{eqnarray}
where $\Delta\varphi_0(\tau)$ is a periodic function.

Next we search for a solution of Eq. (\ref{RSJ1}) in the form $\varphi(\tau)=\varphi_0(\theta(\tau))$,
where $\theta(\tau)$ is an unknown phase satisfying the equation
\begin{eqnarray}
\dot\theta=1+\frac{j+\sin\nu_0\theta}{\nu_0^2}\zeta(\tau).
\label{RSJ2}
\end{eqnarray}
This equation provides a convenient starting point for the perturbation theory in the noise $\zeta(\tau)$.
To the second order in $\zeta(\tau)$ its solution reads
\begin{eqnarray}
\theta = \tau +\int_{0}^\tau d\tau'\frac{j+\sin\nu_0\tau'}{\nu_0^2}\zeta(\tau')
\hspace{2.9cm}
\nonumber\\
+\, \int_{0}^\tau d\tau' \int_{0}^{\tau'} d\tau''\, \frac{\cos\nu_0\tau'(j+\sin\nu_0\tau'')}{\nu_0^3}\zeta(\tau')\zeta(\tau'').
\label{theta}
\end{eqnarray}

Before we turn to the derivation of zero frequency cumulants of the voltage, let us
discuss the time averaging procedure.
Since $\Delta\varphi_0(\tau)$ is a periodic function of time,
it varies within certain finite limits.
Therefore the time derivative $(d/d\tau)\Delta\varphi_0(\theta(\tau))$
vanishes after time averaging for any realization of the noise $\zeta(\tau)$. That means, in turn, that
we can replace $\varphi(\tau)\equiv\varphi_0(\theta(\tau))\to \nu_0\theta(\tau)$ as long as we are interested
only in zero frequency cumulants of the voltage fluctuations.

Thus, deriving $\nu_0\dot\theta$ from Eq. (\ref{theta}) and
averaging it over the realizations of the noise $\zeta(\tau)$ as well as over time,
we find the first cumulant ${\cal C}_1$
\begin{eqnarray}
{\cal C}_1= \nu_0 - \frac{1}{2\nu_0^2}\int_0^\infty d\tau\, {\cal S}_\zeta(\tau)\sin\nu_0\tau,
\label{C1high}
\end{eqnarray}
where ${\cal S}_\zeta(\tau)=\langle\zeta(\tau)\zeta(0)\rangle$ is the pair correlator of dimensionless noise.

Next we evaluate the second cumulant (noise) in the lowest non-vanishing order in $\zeta(\tau)$,
namely in the order $\zeta^2(\tau)$.
We reproduce the known result \cite{Likharev1972,Likharev}
\begin{eqnarray}
{\cal C}_2 =\frac{\nu_0}{2\pi}\int d\tau\int_{-\pi/\nu_0}^{\pi/\nu_0} ds
\langle \delta v(s+\tau/2)\delta v(s-\tau/2) \rangle
\nonumber\\ 
= \int d\tau\,{\cal S}_\zeta(\tau)\left(
\frac{j^2}{\nu_0^2} + \frac{\cos\nu_0\tau}{2\nu_0^2}
\right)
= \frac{j^2S(0)}{\nu_0^2}+\frac{S(\nu_0)}{2\nu_0^2}.
\label{C2high}
\end{eqnarray}

Finally we turn to the third cumulant ${\cal C}_3$.
Here we use the following definition of ${\cal C}_3$
\begin{eqnarray}
{\cal C}_3=\nu_0^3\lim_{t\to\infty} \frac{\nu_0}{2\pi}
\int_{t-\pi/\nu_0}^{t+\pi/\nu_0} \frac{d\tau}{\tau}
\left\langle \left(\theta(\tau)-\langle\theta\rangle\right)^3 \right\rangle.
\end{eqnarray}
One can demonstrate that it is equivalent to the original
formula (\ref{C3def}).
Since the noise $\zeta(\tau)$
is Gaussian, the first non-vanishing contribution to ${\cal C}_3$
comes from the terms $\propto\zeta^4(\tau)$. Making use of the explicit
expression for $\theta$ (\ref{theta}), we arrive at the result
\begin{eqnarray}
{\cal C}_3 &=& \frac{6}{\nu_0^4}\lim_{t\to \infty}\frac{\nu_0}{2\pi}
\int_{t-\pi/\nu_0}^{t+\pi/\nu_0} d\tau
\nonumber\\ && \times\,
\frac{1}{\tau} \int_0^\tau d\tau_1 \int_0^\tau d\tau_2 \int_0^\tau d\tau_3 \int_0^{\tau_3} d\tau_4
\nonumber\\ && \times\,
(j+\sin\nu_0\tau_1) {\cal S}_\zeta(\tau_1-\tau_3)\cos\nu_0\tau_3
\nonumber\\ && \times\,
(j+\sin\nu_0\tau_2){\cal S}_\zeta(\tau_2-\tau_4)(j+\sin\nu_0\tau_4).
\end{eqnarray}
Evaluating the integral we arrive at the main result of this section
\begin{eqnarray}
{\cal C}_3&=& -\frac{3}{8}\,\frac{8j^2S(0)\big(S(0)+S(\nu_0)\big)+S^2(\nu_0)}{\nu_0^5}
\nonumber\\ &&
+\, \frac{3}{4}\,\frac{2j^2S(0)+S(\nu_0)}{\nu_0^4}\,\frac{dS(\nu_0)}{d\nu_0}.
\label{C3high}
\end{eqnarray}
We observe that ${\cal C}_3$ is a sum of two terms. The first one, given by
the first line of Eq. (\ref{C3high}), is always negative. The second contribution
is proportional to the derivative of the noise spectrum and may have any sign.

In a trivial case of white noise
Eqs. (\ref{C1high},\ref{C2high},\ref{C3high}) reduce respectively
to Eqs. (\ref{vax},\ref{likharev},\ref{C3white}). For further illustration
let us again assume that the noise spectrum contains a narrow Lorentzian line and defined by Eq. (\ref{AC})
with $\delta\ll \gamma\omega_0^2/a$.
In the limit of high frequency, $\omega_0\gg 1$, and in the vicinity
of the first Shapiro step, i.e. at $j$ close to $\omega_0$, the cumulants
(\ref{C1high},\ref{C2high},\ref{C3high}) take the form:
\begin{eqnarray}
{\cal C}_1 &=& j -\frac{1}{2j}+\frac{a}{4\omega_0^2}\frac{j-\omega_0}{(j-\omega_0)^2+\delta^2},
\label{c1}
\\
{\cal C}_2 &=& 2\gamma + \frac{a\delta}{2\omega_0^2}\frac{1}{(j-\omega_0)^2+\delta^2},
\label{c2}
\\
{\cal C}_3 &=& \frac{3\gamma a\delta}{\omega_0^2}\frac{ \omega_0-j}{[(j-\omega_0)^2+\delta^2]^2}.
\label{c3}
\end{eqnarray}
Eq. (\ref{c1}) illustrates the formation of the Shapiro step on the I-V curve
at $j=\omega_0$, while the voltage noise (\ref{c2}) reproduces the Lorentzian line
of the input noise spectrum (\ref{AC}). The third cumulant (\ref{c3}) changes its
sign at $j=\omega_0$ and in this regime can be related to the second one by means of a simple
formula ${\cal C}_3 = 6\gamma\,\partial{\cal C}_2/\partial j$.

\section{Discussion and Summary}

We have demonstrated that the intrinsic non-linearity of a 
resistively shunted Josephson junction converts Gaussian
current noise into non-Gaussian voltage noise. The non-Gaussian
properties of the latter are strongest at low temperature
and at bias current below or slightly above $I_c$. It is in this
range of parameters where the non-linearity of the RSJ is most important. It leads,
in particular, to the non-linear voltage-current
characteristics with well resolved critical current.

We have
shown that well below $I_c$ the statistics of voltage fluctuations is
Poissonian and the fluctuations are caused by random phase slip
events. In contrast, at high bias current, $I\gg I_c$, the
statistics becomes Gaussian. In this case the
non-linear properties of the RSJ can be ignored, and it effectively
operates as a linear resistor.

The transition between the two limiting cases turns out to be
non-trivial. With the decrease of the bias current the statistics
gradually deviates from a Gaussian one, and the third and higher
order cumulants of the voltage noise grow. Close to the critical
current, namely for $|I-I_c|\lesssim (2eT/\hbar I_c)^{2/3}$,  the
statistics of voltage fluctuations strongly deviates from both
Gaussian and Poissonian. In this regime the phase slips overlap in
time and are therefore poorly defined, while, at the same time, the
non-linear effects are very strong. Under this conditions the
voltage cumulants of order higher than 3 strongly oscillate
with the bias current. Finally, at $I\lesssim I_c-(2eT/\hbar
I_c)^{2/3}$ the statistics becomes Poissonian. These unusual
properties of a RSJ open up an interesting opportunity of tuning
the statistics of voltage noise by changing the bias current.
Sweeping the current from $I\ll I_c$ to $I\gg I_c$ one can
basically cover all possible statistical distributions.

The general properties of the voltage noise statistics, which we
have just outlined, are found for a weakly frequency-dependent input
current noise spectrum. These properties change if the noise
spectrum contains narrow lines. The latter situation corresponds
to an $ac$-biased RSJ which is known to have Shapiro steps on the
I-V curve. It is also known \cite{Likharev,Barone} that the shape
of every Shapiro step basically repeats that of the I-V curve of
the RSJ in the absence of an external $ac$-signal. We have shown
that the same is true for the third cumulant of the voltage noise,
i.e., the current dependence of ${\cal C}_3$ close to a Shapiro
step qualitatively repeats the dependence  ${\cal C}_3(j)$ of a 
$dc$-biased RSJ, see Fig. \ref{fig:C1C2C3lcolored}. In particular,
depending on the parameters, ${\cal C}_3$ may strongly increase
and change its sign in the vicinity of a Shapiro step. This means,
in turn, that under sufficiently strong microwave irradiation the
statistics of voltage fluctuations in a RSJ may significantly
deviate from Gaussian one even at high bias current $I\gg I_c$
provided the closest Shapiro step on the I-V curve is well
resolved. We expect that not only the third, but all higher order
cumulants should oscillate in the vicinity of the step in the same
way as they do in a $dc$-biased RSJ close to $I_c$.

In summary,
we have investigated the statistics of voltage fluctuations in a resistively
shunted Josephson junction. For white input current noise we have analized
the FCS of these fluctuations  numerically.
In addition we have derived approximate analytical expressions for the
FCS in the limit of strong white noise ($\gamma\gg 1$) for arbitrary bias currents,
as well as, in the limit of weak noise ($\gamma\ll 1$),
for currents below the critical one ($(1-j)^{3/2}\gtrsim\gamma$) and above it
($(j-1)^{3/2}\gtrsim\gamma$). In the intermediate range of currents
($|j-1|^{3/2}\lesssim\gamma$) we observed
the transition from high bias Gaussian voltage fluctuations to low bias
Poissonian statistics of phase slips. In this range the voltage
cumulants of higher than second order oscillate with the bias current, and
the voltage noise may be strongly non-Gaussian.

We have also derived approximate analytical expressions for the third cumulant ${\cal C}_3$ of the
voltage in case of coloured input noise. We have considered two limits: (i) the limit the strong
input noise, and (ii) the limit of weak input noise and bias current higher than the
Josephson critical current.
We have demonstrated that all cumulants ${\cal C}_1,{\cal C}_2,{\cal C}_3$ are sensitive
to high frequency input noise. We have also
shown that ${\cal C}_3$ is enhanced in the vicinity of a Shapiro step caused by
an bias current, and may even change sign
if the $ac$ amplitude is sufficiently large.

\section{Acknowledgments}

We would like to thank Andrei Timofeev and Jukka Pekola for stimulating discussions.
This work has been supported by Strategic International Cooperative Program
of the Japan Science and Technology Agency (JST) and
by the German Science Foundation (DFG).

\end{document}